\documentstyle[epsfig,12pt]{article}

\def\Journal#1#2#3#4{{#1} {\bf #2}, #3 (#4)}

\def\NPA{{Nucl. Phys.} A}
\def\NPB{{Nucl. Phys.} B}
\def\PLB{{Phys. Lett.}  B}

\def\PRL{Phys. Rev. Lett.}
\def\PRC{{Phys. Rev.} C}
\def\PRD{{Phys. Rev.} D}

\newfam\BMath
\font\BMathL=cmmib10 
\font\BMathl=cmmib7
\font\BMathm=cmmib5
\textfont\BMath=\BMathL \scriptfont\BMath=\BMathl
\scriptscriptfont\BMath=\BMathm

\def\P{{\fam\BMath p}}

\def\a{\alpha}

\def\e{\epsilon}

\def\q{\theta}
\def\m{\mu}
\def\n{\nu}

\def\t{\tau}

\def\intps{\int \frac{d^3 \P}{(2\pi)^3}}

\def\exp{\mbox{\rm exp}}

\def\ra{\rightarrow}

\def\lra{\longrightarrow}
\def\llra{\longleftrightarrow}

\def\del{\mbox{$\partial$}}

\def\be{\begin{equation}}
\def\ee{\end{equation}}
\def\bea{\begin{eqnarray}}
\def\eea{\end{eqnarray}}
\def\eref#1{Eq.~(\ref{#1})}
\def\ederef#1#2{Eqs.~(\ref{#1}) and (\ref{#2})}

\def\fref#1{Fig.~\ref{#1}}

\newcommand{\ncom}{\newcommand}
\ncom{\vo}[1]{{\fam\BMath #1}}
\ncom{\vt}[2]{({\fam\BMath #1}+{\fam\BMath #2})}
\ncom{\vmo}[1]{\vert{\fam\BMath #1}\vert}
\ncom{\vmt}[2]{\vert\mbox{\bf #1}+\mbox{\bf #2}\vert}
\ncom{\lan}{\langle}
\ncom{\ran}{\rangle}
\ncom\nonum{\nonumber \\}
\ncom\fx{\!\!\!\!}
\ncom\gsim{\mbox{\raisebox{-0.6ex}{\ $\stackrel {>}{\sim}$\ }}}
\ncom\lsim{\mbox{\raisebox{-0.6ex}{\ $\stackrel {<}{\sim}$\ }}}

\ncom{\half}{{1\over 2}}
\ncom{\third}{{1\over 3}}
\ncom{\fourth}{{1\over 4}}
\ncom{\fifth}{{1\over 5}}
\ncom{\sixth}{{1\over 6}}

\ncom\Tg{T_{eq\; g}}
\ncom\Tq{T_{eq\; q}}
\ncom\qg{\q_g}
\ncom\qq{\q_q}

\value{textfraction}

\font\lbigbf=cmbx10 scaled 1200

\begin{document}

\begin{flushright}
\footnotesize \sffamily LPTHE-Orsay \ 96/69
\end{flushright}

\begin{centering}
{\lbigbf EQUILIBRATION IN HEAVY ION COLLISIONS \\ AT LHC AND AT RHIC}

\vspace{0.25cm}

{S.M.H. Wong}

\em \footnote{Laboratoire associ\'e au Centre National de la
Recherche Scientifique}LPTHE, Universit\'e de Paris XI, B\^atiment 211,
F-91405 Orsay, France

\end{centering}

\begin{abstract}

We consider the evolution of a parton plasma created in
Au+Au collisions at LHC and at RHIC energies. Using Boltzmann
equation, relaxation time approximation and perturbative 
QCD, we show the physics of both thermal and chemical
equilibration in a transparent manner. In particular, 
we show inelastic processes are, contrary to common
assumption, more important than elastic processes,
the state of equilibration of the system can compensate 
for some powers of $\a_s$ for the purpose of equilibration,
the two-stage equilibration scenario is, barring any unknown 
non-perturbative effects, inevitable and is an intrinsic 
feature of perturbative QCD, and gluon multiplication is 
the leading process for entropy generation. 

\end{abstract}
\vspace{-0.5cm}

\section{Introduction}
\label{sec:intro}

Essential questions to ask in heavy ion collisions,
especially at RHIC and at LHC, assuming a gas of weakly
interacting partons can be formed in the central region,
is how fast will this quark and gluon system approaches
equilibrium if the expansion is not too rapid for the 
interactions. If indeed this parton gas can approach
equilibrium, will it be able to end up as a quark-gluon
plasma? And what is the degree of equilibration can
one reasonably expect when the phase transition sets in?
In short, is equilibration fast and how fast?

A number of previous works have already attempted at 
providing answers to these and other related questions of 
equilibration. However, due to the fact that non-equilibrium 
problems are difficult, they addressed either 
only thermalization \cite{baym,gav&etal,shury1,
heis&wang} or only parton chemical equilibration 
\cite{biro&etal1,shury&xion}. Furthermore,
previous attempts at the thermalization problem have only
been done in a heuristic manner without really using
QCD interactions. With the exception of the parton cascade
model (PCM) \cite{geig}, which is based on present knowledge 
of perturbative QCD and some very involved computations,
{\it only then} it is able to consider both thermalization 
and chemical equilibration simultaneously as it should be 
and as it happens in the collisions. In this talk, we
present a relatively simple way to do this and hence
keeping things simple and clear so that the physics 
becomes transparent. 

A much used assumption and starting point in the studies 
of the physics of heavy ion collisions is kinetic 
equilibration is rapid $\lsim 1.0$ fm/c and hydrodynamics
expansion is well underway. PCM has shown that such short
rapid thermalization is too optimistic. In the following,
we will consider the evolution of a parton plasma and
show that complete kinetic
and chemical equilibration are slow, although interactions
can indeed dominate over the expansion, and full equilibration
cannot be attained before the phase transition assumed to
be at $T_c \sim 200$ MeV. We also show that inelastic 
processes are more important than elastic ones in 
equilibration contradicting the common untested assumption 
of the contrary.

\section{To Determine the Evolution of a QCD Plasma}
\label{sec:we}

To study the evolution of a plasma of quarks 
and gluons, it is sufficient to know the particle
distributions. For this purpose, we use the set of rather 
standard assumptions for relativistic heavy ion collisions 
and Baym's form of the Boltzmann equation \cite{baym}
\be \Big ({{\del f(p_\perp, p_z,\t)} \over {\del \t}} \Big )
    \Big |_{p_z \t}=C(p_\perp,p_z,\t)  \; .
\label{eq:baymeq}
\ee
where $\t=\sqrt{t^2-z^2}$, with the collision terms on the right
hand side approximated by the relaxation time approximation
\be C(\vo p,\t)=-{{f(\vo p,\t)-f_{eq}(\vo p,\t)} \over \q(\t)}
\label{eq:Crlx}
\ee
where $f_{eq}$ is the equilibrium distribution 
$f_{eq}=1/(\exp {(p/T_{eq})}\mp 1)$ and $\q(\t)$ is the 
collision time. Whether the plasma equilibrates or not and 
how fast does it equilibrate depends very much on $\q(\t)$ 
\cite{heis&wang,wong}. 

With the combination of \eref{eq:baymeq} and \eref{eq:Crlx},
one can already write down a solution to \eref{eq:baymeq},
which depends, however, on the two numerical parameters 
$T_{eq}$ and $\theta$ that need to be determined from QCD.
We use the simplest QCD interactions at the
tree level for the collision terms for this purpose. 
They are
\be gg \llra ggg \; \; , \; \; \; gg \llra gg \; , 
\label{eq:ggi}
\ee 
\be gg \llra q\bar q \; \; , \; \; \; g q \llra g q \; \; , 
    \; \; \; g\bar q \llra g\bar q \; ,
\label{eq:gqi}
\ee
\be q\bar q \llra q\bar q \; \; , \; \; \; qq \llra qq \; \;  , 
    \; \; \; \bar q \bar q \llra \bar q \bar q \; .
\label{eq:qqi}
\ee
Here to keep things simple and for our purpose, it is
sufficient to include the simplest two leading inelastic 
processes\footnote{Here we assign a single chemical potential, 
$\mu_q$, to all $n_f$ flavours of fermions so flavour changing
interactions in the first interactions of \eref{eq:qqi} is
not considered as inelastic.} the first term of 
\ederef{eq:ggi}{eq:gqi} and all the binary elastic processes. 
Quarks and gluons will be treated as different particle 
species and not as generic partons so that their respective
distributions, $f_g$ and $f_q$, are governed by different 
Boltzmann equations as they should be and depend on 
different collision times, $\qg$ and $\qq$, and different
equilibrium temperatures, $\Tg$ and $\Tq$. 

Using perturbative QCD and suitably infrared regularized
the matrix elements by medium effects, one can construct 
the collision terms for gluons $C_g$ and for quarks $C_q$
semi-classically in the usual way \cite{wong}.
For soft gluon emissions, Landau-Pomeranchuk-Migdal effect 
\cite{gyul&etal} has to be incorporated due to multiple 
scatterings in the medium \cite{biro&etal1,wong}. 

With the real collision terms from QCD, one can construct
two equations for each particle species to solve for the
two time-dependent unknowns in the particle distribution.
We choose the following rate equations. 

\noindent 1) Energy density rate in an one-dimensional 
expanding system
\be {{d \e_i} \over {d \t}}+{{\e_i + p_{L\; i}} \over \t} 
    = -{{\e_i-\e_{eq\; i}} \over \q_i} 
    = \n_i \intps \; p \; C_i (p_\perp,p_z,\t) \; ,
\label{eq:e_trans} 
\ee 

\noindent 2) Collision entropy density rate 
\be  \Big ( {{d s_i} \over {d \t}} \Big )_{coll} 
     = -\n_i \intps \; C_i (p_\perp,p_z,\t) 
     \ln \Big ({{f_i} \over {1 \pm f_i}} \Big )   
     = \n_i \intps {{f_i -f_{eq\; i}} \over \q_i}
     \ln \Big ({{f_i} \over {1 \pm f_i}} \Big )  \; ,
\label{eq:s_rate1} 
\ee
where $i=g,q,\bar q$, $p_L$ is the longitudinal pressure
defined later in \eref{eq:press} and $\n_i$ is the 
multiplicity, for gluons $n_g =2\times 8$ and for quarks 
$n_q =2\times 3\times n_f$. These equations are constructed
from \ederef{eq:baymeq}{eq:Crlx} and $C_i$'s are now understood to
be the real QCD collision terms. 
From them, $\q_i$'s and $T_{eq\; i}$'s can be determined. 
The collision entropy equations, in fact, allow one to break
down the entropy generation process due to collisions into
each of its contributing elements and find out which 
processes are more important for equilibration. We will
show this later on in Sect. \ref{sec:fin}.

\section{The Approach to Equilibrium}
\label{sec:fin}

We take the initial conditions at the isotropic moment
$\t_0$ from HIJING \cite{gyu1} results for Au+Au collisions 
at RHIC and at LHC \cite{biro&etal1,wong}. The
various equations described in Sect. \ref{sec:we} are solved
numerically and the various collective variables can be
calculated. The numerical details and parameters for solving 
the equations can be found in \cite{wong}. Here we 
concentrate on the results and the physics of equilibration.

As mentioned in the introduction, there are two types of 
equilibration: chemical and thermal, which happens 
simultaneously in heavy ion collisions. 
To check the parton composition in the plasma,
it is common to use the concept of fugacity defined as
$l_i =\exp (\m_i/T)$ so that the kinetically equilibrated
distributions can be written as 
$f_i=1/(\exp (p/T_i) l^{-1}_i\mp 1)$. Since chemical
equilibration is essentially $\m_i \ra 0$ starting from some 
initial negative values so that in terms of fugacity, 
$l_i$ approaches $1.0$. Clearly $l_i$'s as well as $T_i's$
exist only when local kinetic equilibrium has been achieved
which is not the case most of the time during the evolution 
of the parton gas. Nevertheless, one can estimate these
quantities from the energy densities $\e_i$'s and number
densities $n_i$'s in the usual way \cite{wong}.

\begin{figure}
\centerline{\epsfig{file=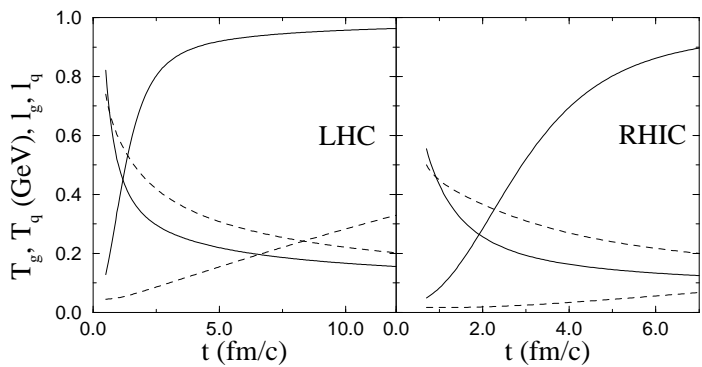,width=3.80in}}
\caption{The evolution of the species temperatures $T_g$ 
and $T_q$ and the fugacities $l_g$ and $l_q$ at LHC and 
at RHIC. The solid (dashed) lines are for gluons (quarks).
The rising (falling) curves are the fugacity (temperature)
estimates.}
\label{fig:temfu}
\end{figure}

These are shown in
\fref{fig:temfu}. The solid lines are for gluons and the dashed
lines are for quarks. The evolution is stopped when the 
temperature estimates all fall below $200$ MeV, the assumed
phase transition temperature.
As can be seen, the gluon fugacities approach $1.0$ much more 
rapidly than those of the quarks both at RHIC and at LHC. 
Gluon fugacities are close to $1.0$ near the end as a result 
but not those of the quarks as in agreement with previous studies
\cite{biro&etal1,shury&xion,geig}. 

\begin{figure}
\centerline{\epsfig{file=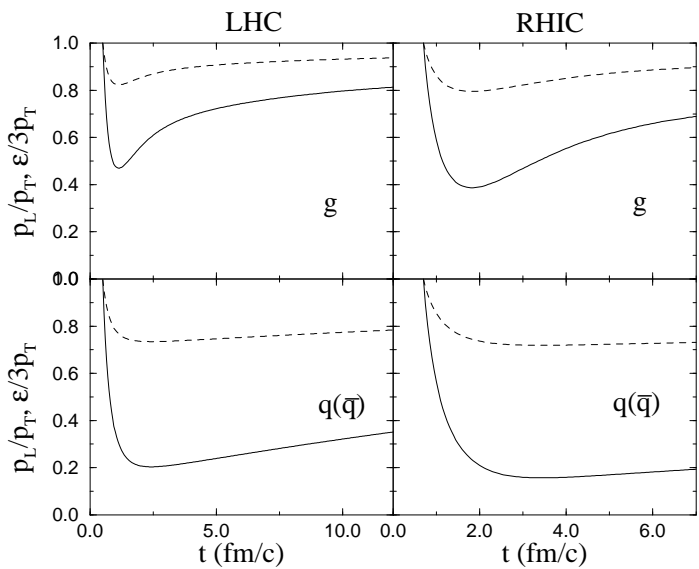,width=3.80in}}
\caption{The plots of the evolution of the ratios of the
longitudinal pressure (solid line) and one-third of the energy 
density (dashed line) to the transverse pressure, $p_L/p_T$ and 
$\e/3p_T$ respectively, for gluons and for quarks at LHC and at 
RHIC.}
\label{fig:press}
\end{figure}

To check for kinetic equilibration, there is not the equivalent
quantity of $l_i$ as for chemical equilibration so instead one
checks the isotropy of momentum distribution by comparing
the ratio of the longitudinal to transverse pressure $p_L/p_T$
and the ratio of a third of the energy density $\e/3 p_T$ 
to the transverse pressure. These pressures are defined by 
\be p_{L,T\; i}(\t)=\n_i \intps {{p_{z,x}^2} \over p} 
    f_i (p_\perp,p_z,\t) \; .
\label{eq:press}
\ee
These quantities are related of course by $\e = 2 p_T +p_L$ 
to the energy density, so isotropy means $p_T = p_L = \e/3$.
These ratios should approach $1.0$ in an equilibrating
parton gas. We have plotted these ratios in \fref{fig:press}
for gluons and for quarks. 

The ratios are $1.0$ initially because we have started
from an isotropic momentum configuration. As the system 
expands, the distributions become anisotropic and reach
maximum anisotropy quickly. The subsequent return towards
isotropy is only progressive and at the end, complete 
isotropy is not fully recovered. For the gluons, one can perhaps
argue an approximate isotropy has been achieved near the
end especially at LHC. For the quarks, they have not
yet passed the half-way mark. To check that, although slow,
these are indeed thermalization behaviours, we can compare 
with the case that the system ends up in free streaming. In 
that case, interactions are not important and can be simulated 
by letting $\qg$ and $\qq$ to $\infty$ and the particle
distributions are described by $f_0$. The corresponding
pressures and ratios work out, as $\t \lra \infty$, to be 
\be p_L/p_T \ra 2 \, \t^2_0/ \t^2 \ra 0 
    \mbox{\hskip 1cm and \hskip 1cm}
    \e/3 \, p_T \ra 2/3
\ee
which are clearly not the behaviours in \fref{fig:press}. 
This provides clear evidence that the interactions
indeed dominate over the expansion. 

\begin{figure}
\centerline{
\begin{tabular}{cc}
\epsfig{file=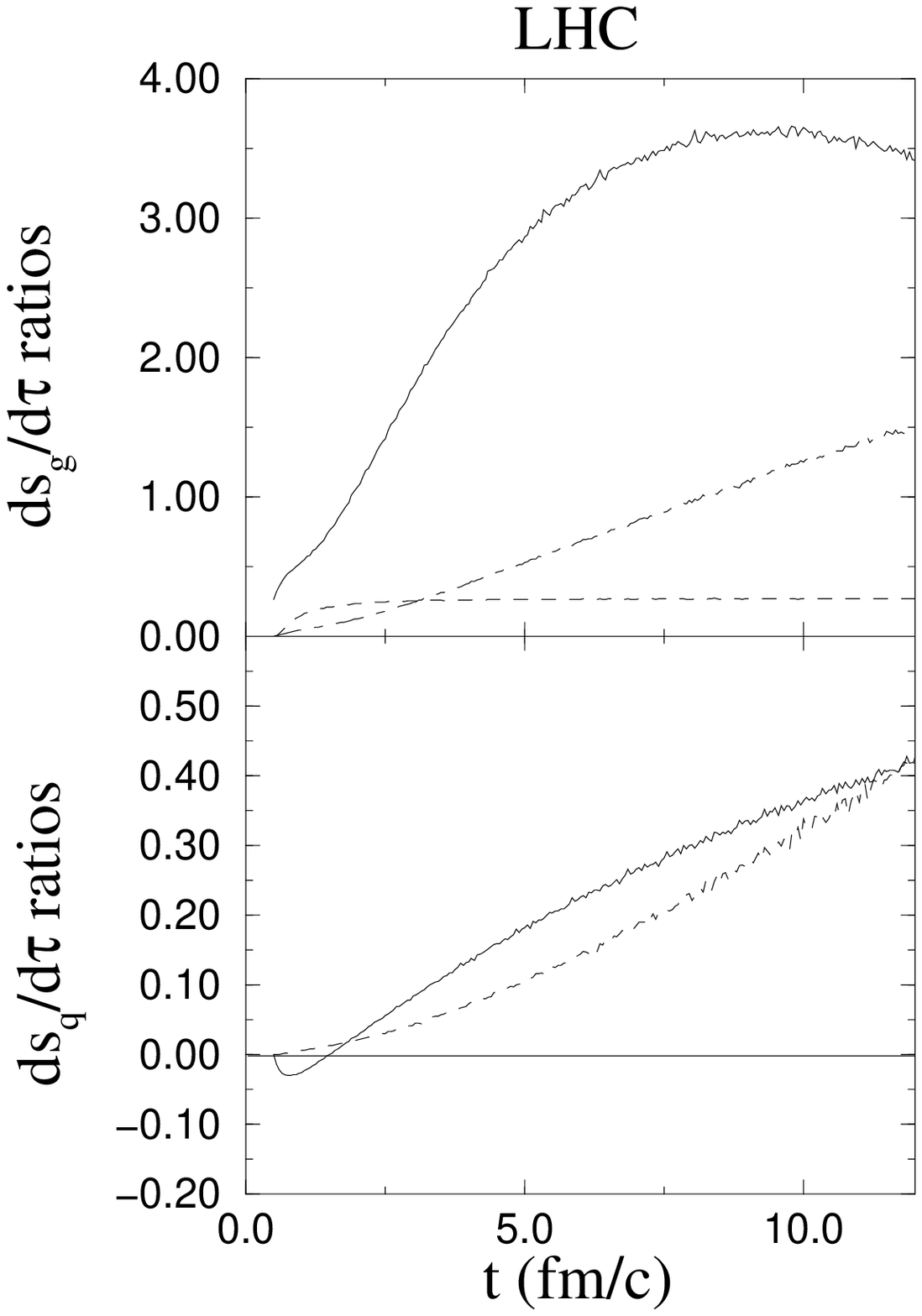,width=2.1in} &
\epsfig{file=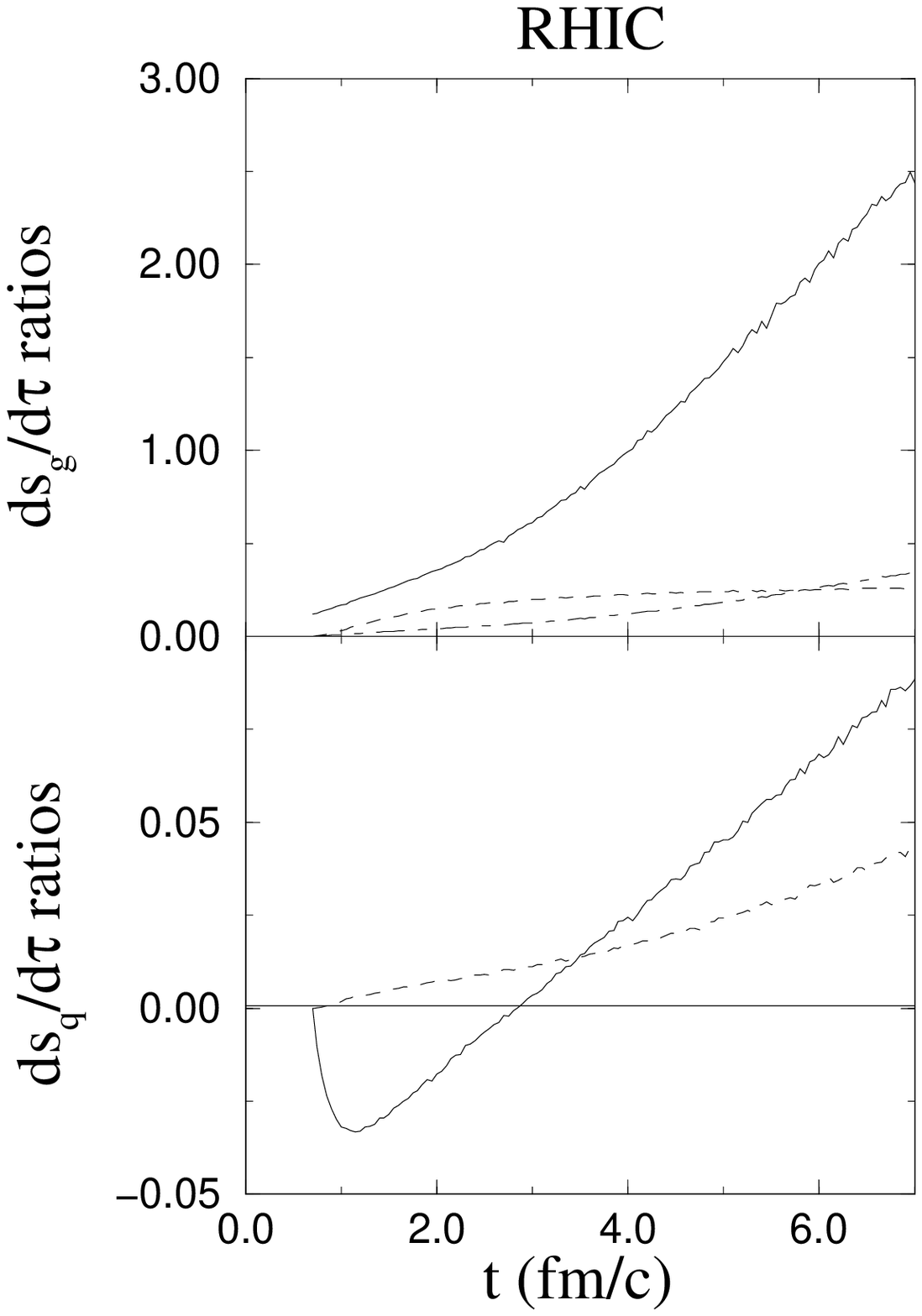,width=2.1in} \\
\end{tabular}
}
\caption{The evolution of the collision entropy density ratios 
of $gg \llra gg$ (dashed), $gg \llra q\bar q$ (solid), and 
$gq \llra gq$ or $g\bar q\llra g\bar q$
(dot-dashed) to $gg\llra ggg$ for gluons (top figures) and 
$gq \llra gq$ or $g\bar q \llra g\bar q$ (solid) and the sum of 
all fermion elastic scatterings to $gg\lra q\bar q$ for quarks 
(bottom figures).}
\label{fig:entropy}
\end{figure}

Finally, we show the dominant processes in the equilibration 
of the plasma. It is common to assume that thermalization
is driven by elastic processes in the studies of the
various physics in heavy ion collisions. Inelastic processes
are relegated to the minor role of essentially only 
for chemical equilibration. To show that this is not true,  
we break down the gluon and quark collision entropy density
rate, $(ds_g/d\t)_{coll}$ and $(ds_q/d\t)_{coll}$ respectively, 
to their contributing elements and plot their ratios. The 
ratios plotted in \fref{fig:entropy} are each process
to gluon multiplication for gluon and to gluon-gluon
conversion into quark-antiquark pair for quark. 

Initially, all the ratios are below $1.0$ in the top figures
i.e. gluon multiplication is dominant, at some point
around $2.0$ fm/c at LHC and $4.0$ fm/c at RHIC.
$gg\llra q\bar q$ rises above one (solid line) and 
overtakes $gg\llra ggg$ as the dominant process. The elastic 
scattering ratios (dashed and dot-dashed lines) remain below 
$1.0$. In the bottom figures, those for quarks, $gg\llra q\bar q$
remains dominant throughout. As can be seen, all the 
curves remain below $1.0$. So for the gluons, gluon
multiplication dominates initially, when this starts to
slow down as equilibrium is getting near and gluons are near
saturation, gluon-gluon annihilation into quark-antiquark
pair and the reverse process become more important because
quarks and antiquarks are still far from completing the
equilibration. For this reason, in the top figures,
gluon-fermion elastic scattering ratios continue to rise
but that of the gluon-gluon elastic scattering is not changing
very much with $\t$. So it is clear that inelastic processes
are more important contrary to common assumption and that
gluon multiplication leads in the production of entropy
as long as they are still far enough from full equilibrium. 

A reason that intuitively elastic processes should be more
important is because of the larger cross-sections. Inelastic
processes such as the radiation of an extra gluon off a
two-body elastic scattering is down by $\a_s$ for instance. 
This reasoning is, however, incorrect. In a medium,
any interaction that can happen can also happen in the reverse
direction, it is not simply the forward or backward reaction 
that enters the collision terms of the Boltzmann equation
\eref{eq:baymeq} but the difference of the two. So it is not
the sizes of the interaction cross-sections which determine 
what processes should be or should not be more important in 
the equilibration of a many body system. 

To summarize, in this talk, we have shown that strictly speaking
equilibration cannot be completed in heavy ion collisions at 
RHIC and at LHC. One can at best consider the system as a fluid 
mixture of an approximately thermalized and chemical equilibrated 
gluon plasma and a still far from equilibrated quark
and antiquark plasma. The usual belief that elastic
processes are responsible for thermalization is flawed.
As we have shown, inelastic processes are even more
important. One has to be very careful in comparing 
interactions in a medium, if it is done by
relying on the sizes of the scattering cross-sections
alone, one is prone to getting the wrong answer.

\section*{Acknowledgements}

The author would like to thank the organizers for this very 
interesting International Seminars and everyone at LPTHE,
Orsay for generous hospitality during his stay there.

\end{document}